\documentclass[preprintnumbers,aps,pre,twocolumn,superscriptaddress,showpacs,amsmath,amssymb]{revtex4}

\usepackage{color}
\usepackage{graphicx}
\usepackage{dcolumn}
\usepackage{bm}

\newcommand{\felastic}{f_{\mathrm{elastic}}}

\newcommand{\sing}{_{\mathrm{s}}}
\newcommand{\nonsing}{_{\mathrm{ns}}}

\newcommand{\NW}{_{\mathrm{NW}}}

\newcommand{\nvec}{\mathbf{n}}
\newcommand{\lvec}{\mathbf{l}}
\newcommand{\mvec}{\mathbf{m}}

\newcommand{\Qvec}{\mathbf{Q}}
\newcommand{\Qij}{Q_{ij}}
\newcommand{\tildeQij}{\tilde Q_{ij}}
\newcommand{\Acal}{\mathcal{A}}

\newcommand{\dd}{\mathrm{d}}

\newcommand{\id}{\mathrm{i}}

\DeclareMathOperator{\Tr}{Tr}

\begin{document}


\title{Scaling of the elastic contribution to the surface free energy of a 
nematic on a sawtoothed substrate}

\author{Jose Manuel Romero-Enrique}
\affiliation{Departamento de F\'\i sica At\'omica, Molecular y Nuclear, Area de F\'\i sica Te\'orica
Universidad de Sevilla, 
Apartado de Correos 1065, 41080 Sevilla, Spain
}%
\author{Chi-Tuong Pham}
\affiliation{Laboratoire d'Informatique pour la M\'ecanique et les Sciences de l'Ing\'enieur, CNRS-UPR 3251, Universit\'e Paris-Sud 11,  BP 133, F-91403 Orsay Cedex, France
}
\author{Pedro Patr\'icio}
\affiliation{Instituto Superior de Engenharia de Lisboa
Rua Conselheiro Em\'\i dio Navarro 1, P-1949-014 Lisboa, Portugal}
\affiliation{Centro de F{\'\i}sica Te\'orica e Computacional, 
Universidade de Lisboa,
Avenida Professor Gama Pinto 2, P-1649-003 Lisboa Codex, Portugal}

\date{\today}

\begin{abstract}

We characterize the elastic contribution to the surface free energy of a
nematic in presence of a sawtooth substrate. Our findings are based on 
numerical minimization of the Landau-de Gennes model and analytical 
calculations on the Frank-Oseen theory. The nucleation of disclination 
lines (characterized by non-half-integer winding numbers) in the wedges and 
apexes of the substrate induces a leading order proportional to $q\ln q$ to
the elastic contribution to the surface free energy density,
$q$ being the wavenumber associated with the substrate periodicity. 

\end{abstract}

\maketitle

\section{Introduction}
Topological defects are ubiquitous in many branches of physics, spanning from
condensed matter physics \cite{Kleman,Mermin} to cosmology 
\cite{Hindmarsh}. They
may emerge in systems with broken continuous symmetry, and their 
presence can introduce essential singularities in the free energy which
lead to infinite-order phase transitions such as the 
Kosterlitz-Thouless transition \cite{Kosterlitz,Nelson}.
Liquid crystalline phases, such as nematics, are prototypical for showing 
topological defects, such as point hedgehogs and disclination lines 
\cite{Oseen,Frank}. Topological arguments show that \emph{within the bulk} and in 3D 
the only stable disclination lines have $\pm 1/2$ winding number \cite{Mermin}. 
The presence of substrates may induce the nucleation of disclinations with 
other winding numbers. Theoretically, non-half-integer disclination lines are
predicted to nucleate in the cusps (wedges and/or apexes) of substrates 
\cite{Barbero_1980,Barbero_1981,Barbero_1982,Poniewierski_2010}, as well as
in sharp boundaries between domains characterised by different anchoring conditions in flat surfaces \cite{Evangelista_1994,Cardoso_1996}.
Furthermore, there is some experimental evidence of the formation of these
unusal non-half-integer disclination lines on surfaces \cite{Dolganov}.   

It is well known that the nematic director field in presence
of structured substrates may be distorted, leading to an elastic 
contribution to the free energy. Since the seminal work by Berreman 
\cite{Berreman,Pgg}, this problem has been extensively studied and generalized
in the literature \cite{Barbero_1980,Barbero_1981,Barbero_1982,Brown_2000,
Kitson_2002,Fukuda,Barbero,Yi_2009,Poniewierski_2010}. 
However, most of these studies 
focussed on smooth substrates or on the effect of isolated cusps. In this paper 
we will consider the effect that the disclination lines nucleating on the
cups of the substrates have on the
elastic contribution to the surface free energy of the nematic on periodic
and cusped structures. In particular, we will consider a sawtoothed substrate
which favors homeotropic anchoring.
Other geometries may also be considered with our formalism, such as
step-like substrates, which have been studied numerically and experimentally
in the context of zenithal bistable switching in nematic devices
\cite{Brown_2000,Parry-Jones1,Parry-Jones2}. 
We will show that the disclination lines
induce a contribution to the elastic contribution to the free energy density
(per unit projected area) which
depends only on the geometric characteristics of the substrate, and that scales
with the wavenumber $q$ associated to the periodicity of the substrate as 
$-q\ln q$.  

The paper is organized as follows. The different 
models for the nematic in the presence of a sawtooth substrate are presented in Section II. We focus in particular on substrates which favor homeotropic anchoring and their consequences. In Section III we
report the numerical and analytical results obtained from these approaches. 
In Section IV we present our conclusions.
 
\section{The model}

\begin{figure}[t] 
\centerline{\includegraphics[width=.95\columnwidth]{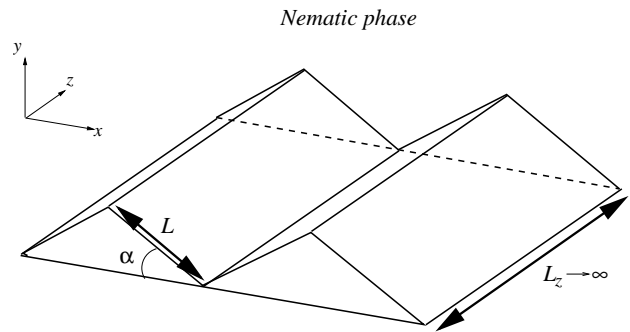}}
\caption 
{Schematic picture of the geometry of the system, characterized by the 
side length $L$ and the angle $\alpha$. 
} 
\label{fig1}
\end{figure}

We 
consider a nematic phase in contact with a sawtooth substrate 
characterized by an angle $\alpha$ and
a side length $L$ (see Fig. \ref{fig1}). At the substrate the nematic molecules
preferentially align homeotropically, i.e.
parallel to the local normal to the substrate. The system is translationally 
invariant along the out-of-plane axis $z$ and periodic along the $x$ axis. 
We will only consider azimuthal distortions for the nematic director $\nvec$, 
which can be parametrized in terms of the angle $\theta$ between the director 
and the $y$ direction as $\nvec=(-\sin\theta,\cos\theta,0)$.  
Out-of-plane or twist deformations may also be important under other 
conditions, as a twist
instability may occur~\cite{Patricio_Telo_Dietrich_2002}, but we 
checked numerically they are not relevant for our choice of parameters as free-energy minimization always 
leads to azimuthal distorted textures.  
Finally we impose that, far from the substrate, the bulk nematic phase  
orients homogeneously either along $x$ (the $N^\parallel$ texture) or the 
$y$ direction (the $N^\perp$ texture), which  
are the only relevant situations allowed by symmetry considerations.  
We study the orientational ordering of this system within two different
models: the Landau-de Gennes (LdG) model and the Frank-Oseen (FO) model. 

\subsection{The Landau-de Gennes model}

In the LdG model, both isotropic and nematic phases can 
be locally represented by a traceless, symmetric order-parameter tensor with 
components $\Qij$, which can be represented as 
$\Qij = \frac 3 2 S[n_in_j - \frac 1 3 \delta_{ij}] + \frac 1 2 B
[l_il_j - m_i m_j]$,
where $n_i$ are the Cartesian components of the director field $\nvec$, 
$S$ is the nematic order parameter which measures the orientational ordering 
along the nematic director, 
and
$B$ is the biaxiality parameter, which 
measures the ordering of the molecules on the orientations 
perpendicular to $\nvec$, characterized by the eigenvectors $\lvec$ and 
$\mvec$. 
The LdG free energy can be written as
$\mathcal{F}_{\mathrm{LdG}} = \int_{\mathcal{V}} (\phi_{\mathrm{bulk}} +
\phi_{\mathrm{el}})\,\dd V + \int_{\mathcal{A}} \phi_{\mathrm{surf}}\,\dd s$
where $\phi_{\mathrm{bulk}}$ is the bulk free energy density,
$\phi_{\mathrm{el}}$ is the elastic free energy density, and
$\phi_{\mathrm{surf}}$ is the surface free energy are defined as~\cite{Pgg}: 
\begin{align}
& \phi_{\mathrm{bulk}} = a \Tr \Qvec^2 - b \Tr
\Qvec^3 + c [\Tr \Qvec^2]^2&&\\ & \phi_{\mathrm{el}} =
\frac{L_1}{2}\partial_k \Qij\partial_k \Qij + \frac{L_2}{2} \partial_j \Qij
\partial_k Q_{ik} &&\\ & \phi_{\mathrm{surf}} = - \frac 2 3 w \Tr
[\Qvec\cdot\Qvec_{\mathrm{surf}}]
\end{align}
where $a$ depends linearly on the temperature, $b$ and $c$ are positive 
constants, and $L_1$ and $L_2$ are positive parameters related to the elastic 
constants. If we will rescale all the variables as follows 
\cite{Andrienko_Tasinkevytch_Patricio_etal_2004}:
$\tilde \Qvec=6c\Qvec/b$, 
the positions $\tilde{\mathbf{r}}={\mathbf{r}}/\xi$, where the correlation 
length $\xi$ is 
defined as $\xi^2=8 c (3 L_1+ 2L_2)/b^2$, and 
$\mathcal{\tilde F}_{\mathrm{LdG}} 
= 24^2 c^3 \mathcal{F}_{\mathrm{LdG}}/\xi^3 b^4$, we get that 
$\mathcal{\tilde F}_{\mathrm{LdG}} = \int_{\mathcal{\tilde V}} (\tilde 
\phi_{\mathrm{bulk}} +
\tilde \phi_{\mathrm{el}})\,\dd \tilde V + \int_{\mathcal{\tilde A}} \tilde 
\phi_{\mathrm{surf}}\,\dd \tilde s$, with rescaled free energy densities:
\begin{align}
& \tilde \phi_{\mathrm{bulk}} = \frac 2 3 \tau \Tr \tilde \Qvec^2 - \frac 8 3 \Tr
\tilde \Qvec^3 + \frac 4 9 [\Tr \tilde \Qvec^2]^2&&\\ & \tilde \phi_{\mathrm{el}} = \frac {1}
{3+2\kappa}[\tilde \partial_k \tildeQij\tilde \partial_k \tildeQij + 
\kappa \tilde \partial_j 
\tildeQij
\tilde \partial_k \tilde Q_{ik}] &&\\ & \tilde \phi_{\mathrm{surf}} = - \frac 2 3 \tilde w \Tr
[\tilde \Qvec\cdot\tilde \Qvec_{\mathrm{surf}}]
\end{align}
Here $\tau=24ac/b^2$ is a dimensionless temperature, $\kappa=L_2/L_1$ is an 
elastic dimensionless parameter ($\kappa>-3/2$) and 
$\tilde w=16wc/b^2 \xi $ is the dimensionless anchoring strength. 
Hereafter we will consider these rescaled expressions, so we will drop the 
tilde notation. For $\tau=1$, 
the bulk free-energy density has two minima corresponding to 
$\phi_{\mathrm{bulk}}=0$ for rescaled scalar order parameters $S_\mathrm{I}=0$ 
(isotropic phase) and $S_\mathrm{N}=1$ (nematic phase), so both phases are
at coexistence. It is important to note that the order parameter $S$ in
the coexisting nematic phase is rescaled, so its value in real units is 
$b/6c$, which must be smaller than 1 (typically $\approx 0.4$).
If the elastic parameter $\kappa$ is positive
(negative), the nematic prefers to align parallel (perpendicular) to
a possible nematic-isotropic interface. Finally, $\Qvec_{\mathrm{surf}}$
defines the favored tensor at the substrate. 
We will favor a homeotropic alignment of the nematic 
by setting $\Qvec_{\mathrm{surf}}=(3\boldsymbol{\nu}\otimes
\boldsymbol{\nu}-1)/2$, with $\boldsymbol{\nu}$ the normal vector to the 
substrate, 
establishing a
direct connection to previous papers 
\cite{Sheng_1976,Sheng_1982,Braun_1996,Patricio_Pham_Romero-Enrique}. 
Hereafter we will restrict ourselves to the nematic phase at the 
nematic-isotropic (NI) transition ($\tau=1$), with $\kappa=2$. 

\subsection{The Frank-Oseen model}

The FO model can be considered as an approximation to the LdG model, in which 
we assume that the variations of the nematic order parameter are restricted 
to the neighbourhood of the substrate (with a
width of order of $\xi$) and inside defect cores, and takes the bulk 
value $S$ elsewhere. 
Assuming there are no disclination lines in the bulk, substitution of this ansatz 
into the LdG free-energy functional leads to the following FO approach for 
large $L$ (in units of $\xi$):
\begin{equation}
\mathcal{F}\approx \mathcal{F}_{\mathrm{FO}}\equiv \phi_{\mathrm{bulk}}(S)
\mathcal{V}+\int_{\Acal} \dd s \Sigma(\theta(s)) + 
\frac{K}{2}\int_{\mathcal{V}} \dd V 
|\boldsymbol \nabla \theta|^2 
\label{mfofreeenergy}
\end{equation}
where $\mathcal{V}$ is the total volume of the nematic, 
$\Acal$ is the substrate area in contact with the nematic, 
the elastic constant $K=(9/2)S^2(2+\kappa)/(3+2\kappa)$ and
$\Sigma(\theta)$ is an effective anchoring potential to be 
determined in the following that is due to the nematic order
parameter distorsions close to the substrates.  

\begin{figure}[t]
\centerline{\includegraphics[width=.9\columnwidth]{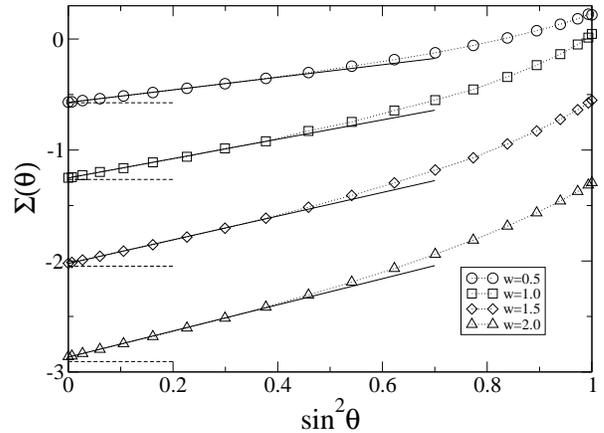}}
\caption{Plot of the effective anchoring potential $\Sigma(\theta)$ as
a function of $\sin^2\theta$, for a layer width $\eta=1.5\xi$ and 
$w=0.5$ (circles), $w=1$ (squares), $w=1.5$
(diamonds) and $w=2$ (triangles). The dashed horizontal 
lines correspond to the values
of $\sigma\NW$ obtained from Eq. (\ref{sigmanw}), and the continuous lines
correspond to the fit to a Rapini-Papoular expression for small values of
the tilt angle. The values of the slopes $\Sigma_0''$ are $0.565_5$ ($w=0.5$),
$0.872_9$ ($w=1$), $1.067_8$ ($w=1.5$) and $1.177_{11}$ ($w=2$). 
\label{fig2}}
\end{figure}

In order to make a quantitative comparison with the results within the LdG 
model, instead of considering an \emph{ad hoc} expression for the anchoring
potential, we will derive it from the LdG model. As it was mentioned above,
the nematic order parameter distortions are assumed to be confined to a layer of width 
$\eta\sim \xi$ close to the substrates. On the other hand, the variations 
of the nematic director field close to the walls are rather small in 
directions parallel to the surfaces for large $L$
(except close to the wedges and apexes). 
So, we may estimate $\Sigma(\theta)$ as the LdG excess free energy per 
unit area of a slab of 
width $\eta$, for which, at the boundary $y=0$, we impose a surface field   
$\phi_{\mathrm{surf}}$, and for $y\ge \eta$ we consider a bulk nematic phase
with an uniform nematic director characterized by a tilt angle $\theta$ with
respect to the $y$ axis. Fig.~\ref{fig2} shows $\Sigma(\theta)$ obtained by
numerical minimization with a conjugate-gradient method for $\eta=1.5\xi$
(similar results are obtained for other values of $\eta$). For each value
of $w$, the minimum value of $\Sigma(\theta)$ corresponds to the homeotropic
alignment $\theta=0$, where it takes the (true) nematic-wall surface tension 
value $\sigma\NW(w)$. For not too large values of $\theta$, $\Sigma(\theta)$
takes a Rapini-Papoular form \cite{Rapini-Papoular} 
$\Sigma(\theta)\sim \sigma\NW + \Sigma_0''(w) 
\sin^2 \theta$. There are deviations for $\theta$ around $\pi/2$, but in 
any case the involved energies are much larger. In the range of values of $w$
we will consider in this paper ($0<w<2$), both $\sigma\NW$ and $\Sigma_0''$ 
are of order of $w$. Indeed $\sigma\NW$ 
can be obtained analytically from minimization 
of the LdG functional in the presence of a flat wall and for homeotropic 
anchoring \cite{Sheng_1976,Sheng_1982}. 
In this situation the equilibrium nematic director field does not show any 
deformation and there is no biaxiality, so the resulting free energy at
NI coexistence can be expressed in terms of the nematic order parameter profile 
$S=S(y)$ as 
$\mathcal{F}_{\mathrm{LdG}} = \Acal \int_0^\infty (S^2-2S^3+S^4+
(S')^2/2)\dd y-w S(0)$.
where $S'(y)=\dd S/\dd y$. 
By introducing a magnetization-like field $m(y)\equiv 2S(y)-1$, 
the LdG functional reduces to the Landau-Ginzburg free-energy 
functional of an Ising model for zero applied magnetic field 
in presence of a flat wall. This problem has been extensively studied 
in the literature \cite{Binder_Hohenberg_1972,Cahn_1977}, leading to 
a nematic order parameter profile $S(y)=(1+g\exp(-\sqrt{2}y))^{-1}$, 
where $g$ is obtained from the boundary condition $S'(0)=-w$.
The resulting expression for $\sigma\NW$ is:
\begin{equation}
\sigma\NW=\frac
{\sqrt{3}(g+3)g^2}{6(1+g)^3}-\frac{w}{1+g}
\label{sigmanw}
\end{equation}
where $g=-(1+1/\sqrt{2}w)+\sqrt{(1+1/\sqrt{2}w)^2-1}$. 

In order to obtain the azimuthal angle $\theta$ field, we have to
minimize the energy functional given by Eq. (\ref{mfofreeenergy}). In the bulk
, we
thus have to solve the Laplace equation $\nabla^2 \theta=0$ with
appropriate anchoring conditions at the boundaries. Let
$\boldsymbol{\nu}=(-\sin\theta_0, \cos\theta_0,0)$ now be the local normal
to the substrate. Using the approximation $\Sigma(\theta)\sim
\sigma\NW + \Sigma_0''(w) \sin^2 (\theta - \theta_0)$ found previously, the 
nematic director field then satisfies approximately that $K
\boldsymbol{\nu}\cdot \boldsymbol{\nabla}\theta + \Sigma_0''
\sin(2(\theta-\theta_0))=0$. Note that we may assume strong
homeotropic anchoring conditions when $L$ is large compared to the
extrapolation length $K/2\Sigma_0''\sim K/w$, or equivalently $wL\gg
1$.  We can justify this assumption by considering the following
rescaling: $\mathbf{r}^*=\mathbf{r}/L$,
$\theta^*(\mathbf{r}^*)=\theta(\mathbf{r})$ and
$\mathcal{F}_{\mathrm{FO}}^*=\mathcal{F}_{\mathrm{FO}}/L$. In this
rescaled description, we must minimize the FO free energy
(\ref{mfofreeenergy}) in the rescaled domain subject to an effective
anchoring potential $\Sigma^*(\theta^*)=L\Sigma(\theta)$. This means
that the dependence on the size of the system can be absorbed into the
coefficients of the anchoring potential: $\sigma\NW$ (which does not
affect the anchoring conditions) and $\Sigma_0''$ (which are rescaled
by a factor of $L$). Strong anchoring condition is then satisfied when
the rescaled extrapolation length $K/2\Sigma_0''L\ll 1$, in agreement
with our previous estimate.

\begin{figure}[t]
\centerline{\includegraphics[width=.95\columnwidth]{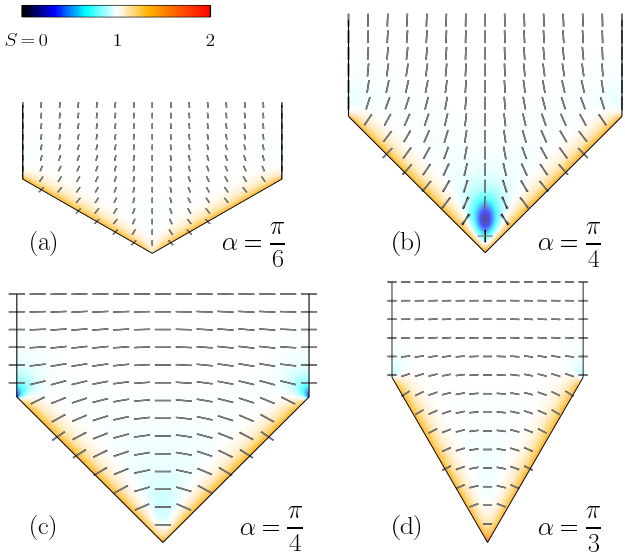}}
\caption{(Color online) 
Contour maps of the nematic order parameter $S$ (blue (dark grey)
for small $S$, white for the bulk value $S=1$, and orange (light grey) 
for higher $S$) and the nematic director field $\nvec$ (grey segments) for
the equilibrium textures obtained from minimization of the LdG model for 
$L=16$ and $w=1.0$: (a) $\alpha=\pi/6$ ($N^\perp$), (b) $\alpha=\pi/4$ 
($N^\perp$), (c) $\alpha=\pi/4$ ($N^\parallel$) and (d) $\alpha=\pi/3$ 
($N^\parallel$). The singular nematic director field  
from the FO approach (Eqs. (\ref{thetasingperp}) and (\ref{thetasingpar})) are
also shown for comparison (black segments). Note that they
are indistinguishible almost everywhere. \label{fig3}}
\end{figure}

\section{Results}
\subsection {Numerical results within the LdG model}
We now turn to the evaluation of the distortion contribution in the LdG model
We consider values of $\alpha$ between $0$ and $\pi/2$, $8\le L \le 96$
and $0\le w\le 2$. In order to get the orientational ordering, we numerically 
minimize the LdG free-energy functional by using a conjugate-gradient method. 
The numerical discretization of the continuum problem is performed with a 
finite element method combined with adaptive meshing in order to resolve the 
different length scales that may emerge in the problem 
\cite{Patricio_Tasinkevych_Telo_2002}. The numerical procedure is similar 
to that used to study the wetting transition by the nematic phase in this 
geometry \cite{Patricio_Pham_Romero-Enrique}. The numerical minimization 
of the LdG functional $\mathcal{F}_{\mathrm{LdG}}$ shows that the $N^\perp$ 
texture has lower free energy if $\alpha<\pi/4$ owing to lesser distortion. 
Conversely, the $N^\parallel$ texture has lower free energy for $\alpha>\pi/4$.
These results are in agreement with previous studies in the literature 
\cite{Barbero_1980,Barbero_1981,Barbero_1982,Poniewierski_2010}.
As a consequence, we observe bistability in a range of values of $\alpha$ 
around $\pi/4$. Fig. \ref{fig3} displays some 
typical textures. We see that on the substrates the nematic orientation
is preferentially homeotropic. Inspection of these textures show that in 
general there are no disclinations in bulk, except for large $w$ and $\alpha$
around $\pi/4$, but there is an important distortion on the nematic director
field close to the wedges and apexes. On the other hand, when disclinations are
observed in bulk (see, for example, Fig.~\ref{fig3} (b) or (c)), they are at
a distance of order of $\xi$ from a wedge or apex, which is almost independent
of $L$.  

\begin{figure}[t]
\centerline{\includegraphics[width=.9\columnwidth]{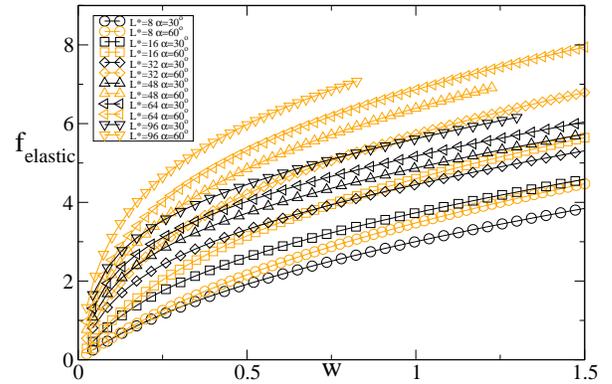}}
\caption{(Color online) 
Plot of  $\felastic$ as a function of the anchoring strength $w$ for
$\alpha=\pi/6$ (black symbols) and $\alpha=\pi/3$ (orange 
(light grey) symbols). We 
considered different cell sizes: $L=8$ (circles), $L=16$ (squares), $L=32$
(diamonds), $L=48$ (triangle up), $L=64$ (triangle left) and $L=96$ (triangle
down). \label{fig4}}
\end{figure}

The analysis of the calculated equilibrium free 
energies confirms that the leading order contribution to the equilibrium 
free energy per unit area is $\sigma\NW$, which is a further confirmation 
that there is strong anchoring on the substrates.
As a consequence, we may obtain the elastic part of the free energy per cell 
(i.e. a slice parallel to the $yz$ plane with width along the $x$ axis equal to 
the substrate period) and unit length in the $z$ direction,
$\felastic$, as $2L (\mathcal{F}_{\mathrm{LdG}}|_{\mathrm{eq}}/\mathcal{A} 
-\sigma\NW)$. Numerical evaluation of $\felastic$ shows a clear dependence 
on $L$ in a broad range of values of $w$ (see Fig.\ref{fig4} for 
$\alpha=\pi/6$ and $\alpha=\pi/3$). 
It is interesting to compare this result to the prediction within 
Berreman's approach for the elastic contribution to a smooth substrate-nematic 
surface free energy \cite{Berreman}. 
Although originally this approach was introduced 
for shallow sinusoidal substrates with strong tangential anchoring,
it can be extended straigthforwardly to the present case
with homeotropic anchoring, either weak or strong. By using the electrostatic
analogy, we may expand the azimuthal angle field $\theta$ as:
\begin{equation}
\theta(x,y)=\sum_{n=0}^\infty M_n\sin (nqx) \exp(-nqy)
\label{expansion}
\end{equation}
where $q$ is the wavenumber associated to the substrate periodicity, and 
the coefficients $M_n$ are chosen to satisfy the imposed boundary conditions.
Substitution of this expression in the FO free energy 
leads to the elastic contribution to the surface free energy which depends 
only on $Aq$, with $A$ the roughness amplitude. For sinusoidal substrates
under moderate or strong anchoring conditions, Berreman's approximation 
$\felastic \approx \pi K (Aq)^2$ is valid if $qA$ is smaller than 1
\cite{Barbero}. In any case, if $A$ scales like $L$, the surface free energy 
should be independent of $L$. In order to explain and characterize the 
anomalous scaling for the elastic free energy obtained within the LdG 
framework, we will now resort to the FO approximation.

\subsection{Results with the FO model}
Disclination lines are known to nucleate on wedges and apexes 
\cite{Barbero_1980,Barbero_1981,Barbero_1982,Poniewierski_2010}. Their
presence introduce a singular contribution to the orientation field $\theta$
which cannot be expanded as Eq. (\ref{expansion}). We may write $\theta(x,y)$ 
as $\theta\sing(x,y)+\theta\nonsing(x,y)$, where $\theta\sing$ is the singular
contribution due to the disclination lines, and $\theta\nonsing$ is the 
non-singular part. Note that Berreman assumed that $\theta_s\equiv 0$. 
The singular $\theta$ field associated to one isolated disclination line 
located at the origin and charaterized by a winding number $I$
takes the form $-I\arctan(x/y)=I\Im(\ln(\id\zeta))$, where 
$\zeta=x+\id y$ and $\Im(\zeta)$ is the imaginary part of $\zeta$. 
In the bulk, the values of the winding number $I$ are restricted to 
half-integer values to avoid discontinuities in the nematic director field. 
For a periodic linear array of disclination lines, we may obtain 
the singular nematic orientation field from a conformal transformation:
\begin{eqnarray} 
\zeta\mapsto \zeta'=\sin\left(\frac{q\zeta}{2}\right)
\label{conformalmapping}
\\=
\sin\left(\frac{qx}{2}\right)
\cosh\left(\frac{qy}{2}\right)
+\id\cos\left(\frac{qx}{2}\right)\sinh\left(\frac{qy}{2}\right)
\nonumber
\end{eqnarray}
where the wavenumber which characterizes the substrate periodicity is
$q=\pi/L\cos\alpha$.
The transformation maps a neighbourhood of any  
point $\zeta_n=n(2L\cos\alpha)$, with $n$ integer, to a neighbourhood of the
origin in the $\zeta'$-complex plane. If we denote $\zeta=\zeta_n+\epsilon$ 
($|\epsilon|$ small), then the transformation leads to a complex
number $\zeta'\approx (-1)^n q\epsilon /2$. 
Thus the orientation field $I\Im(\ln(\id\zeta'))=
-I\arctan(\tan(qx/2)\coth(qy/2))$, which is a 
solution to the Laplace equation, periodic in $x$ with period
$2L\cos\alpha$, and singular at each $\zeta_n$, reduces to the field 
corresponding to an isolated defect of winding number $I$ in the neighbourhood 
of any $\zeta_n$. Focussing on the range $x\in [-L\cos\alpha,L\cos\alpha]$, the
azimuthal angle generated by this function is $\theta=0$ at $x=0$ and $x=\pm L
\cos\alpha$. On the other hand, for $|qy/2|\gg 1$, $\theta \approx 
-\mathrm{sgn}(y) Iqx/2$ for $x\ne \pm L\cos\alpha$, where $\mathrm{sgn}(y)=
y/|y|$ is the sign of $y$. Note that there is a 
jump from $\theta=-\mathrm{sgn}(y)I \pi/2$ to $\theta=
\mathrm{sgn}(y)I \pi/2$ when crossing $x=L\cos\alpha$.  
In the bulk, this solution is 
acceptable if $I$ is integer (note that the nematic state is invariant under
inversion of the nematic director). If $I$ is half-integer, we may add to this 
solution the orientation field $I\arctan(\tan(qx/2))$ due to an array of disclination lines with
winding number $-I$ located at $\zeta_n-\id\infty$,
which in the $x$ interval $(-L\cos\alpha,L\cos\alpha)$ reduces to $Iqx/2$.    
The compound orientation field $I[-\arctan(\tan(qx/2)\coth(qy/2))+
\arctan(\tan(qx/2))]$ has no discontinuity at $x=\pm L\cos\alpha$
for $y>0$ (in fact, it goes to zero as $qy/2\gg 1$), but it jumps from
$I\pi$ to $-I\pi$ when crossing $x=L\cos\alpha$ for $y<0$. 
This is again physically acceptable if $I$ is half-integer. Finally, we 
must mention that there are 
different physically equivalent representations of the same singular field 
associated to a periodic array of disclination lines located at $x=\zeta_n$ as,
for example, the field $I[\arctan(\cot(qx/2)\tanh(qy/2))-\arctan(\cot(qx/2))]$.
This orientation field shows a physically acceptable discontinuity  
at $x=0$ for $y<0$ if $I$ is half-integer.
 
When disclinations lines are located on surfaces, their winding numbers $I$ are 
not constrained to half-integer values \cite{Barbero_1980,Barbero_1981,
Barbero_1982,Poniewierski_2010}, as we may tune the winding numbers in order 
to match the boundary conditions close to any wedge or the apex in the
strong anchoring regime. Alternatively, a Schwartz-Christoffel transformation
may be used \cite{Barbero_1980}.
For the $N^\perp$ texture, we find that the winding numbers $I_1$ and $I_2$
for the disclination lines at the wedge bottom and at the apex top, 
respectively, are $I_1=-\alpha/(\pi/2-\alpha)$ and $I_2=\alpha/(\pi/2+\alpha)$.
For the $N^\parallel$ texture, the topological charges are $I_1=1$ and
$I_2=-(\pi/2-\alpha)/(\pi/2+\alpha)$. The resulting singular orientation
fields, $\theta_s^\perp$ and $\theta_s^\parallel$, can be written in terms
of the wavenumber $q$ and for $x\in (-L\cos\alpha,L\cos\alpha)$ as:
\begin{eqnarray}
\theta_s^\perp&=&\frac{-\alpha}{\frac{\pi}{2}-\alpha}\Bigg(-\arctan
\left[\tan\frac{qx}{2}\coth\frac{qy}{2}\right]\nonumber\\
&+&\arctan
\left[\tan\frac{qx}{2}\right]\Bigg)
\label{thetasingperp}\\
&+&\frac{\alpha}{\frac{\pi}{2}+\alpha}\Bigg(-\arctan
\left[\tan\frac{qx}{2}
\tanh\frac{q(y-L\sin\alpha)}{2}\right]
\nonumber\\
&+&\arctan\left[\tan\frac{qx}{2}\right]\Bigg) 
\nonumber\\
\theta_s^\parallel&=& \frac{\pi}{2}+
\Bigg(-\arctan
\left[\tan\frac{qx}{2}\coth\frac{qy}{2}\right]\nonumber\\
&+&\arctan
\left[\tan\frac{qx}{2}\right]\Bigg)
\label{thetasingpar}\\
&-&\frac{\frac{\pi}{2}-\alpha}{\frac{\pi}{2}+\alpha}\Bigg(-\arctan
\left[\tan\frac{qx}{2}
\tanh\frac{q(y-L\sin\alpha)}{2}\right]
\nonumber\\
&+&\arctan\left[\tan\frac{qx}{2}\right]\Bigg)
\nonumber
\end{eqnarray}
This expression does not show any discontinuity above the substrate.
We checked that these solution approximate quite well the boundary
anchoring conditions (but not exactly), so that
$\theta\nonsing$ may be neglected in most of the cases.
Furthermore, 
the agreement between our \emph{ansatz} for the director orientation field
and the numerical LdG textures is excellent except close to the defect
cores, as can be seen in Fig. \ref{fig3}.  

\begin{figure}[t]
\centerline{\includegraphics[width=.9\columnwidth]{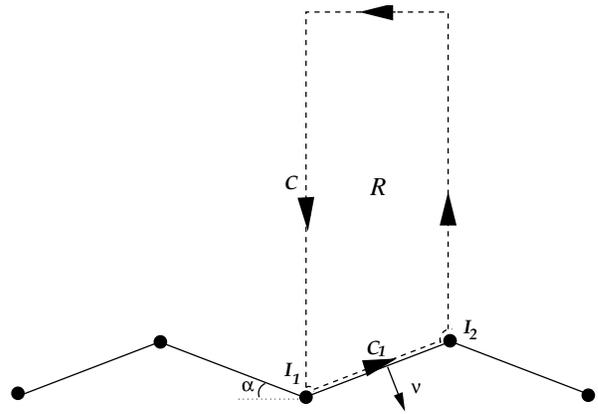}}
\caption{Plot of the integration contour $C$ (dashed line) to get the elastic 
contribution to the free energy in the region $R$ inside the contour. 
The contour is deformed to avoid the cores of the disclination lines with 
wandering numbers $I_1$ and $I_2$ at the wedge 
and apex (filled circles). We highlight the contour $C_1$ with an associated 
outwards normal $\boldsymbol{\nu}$.
\label{fig5}}
\end{figure} 

The elastic energy can be obtained by standard contour integration techniques. 
By symmetry, $\felastic$ can be obtained as twice the contribution of the 
half-cell region $R$ (see Fig.~\ref{fig5}). By using the identity
$\boldsymbol{\nabla}\cdot (\theta \boldsymbol{\nabla}\theta)=
|\boldsymbol{\nabla}\theta|^2+
\theta \nabla^2 \theta$, and using the divergence theorem and the fact
that $\nabla^2 \theta=0$ in $R$, we obtain that $\felastic$ can be expressed
in terms of an integral over the contour $C$ of $R$:
\begin{equation}
\felastic=K\oint_C  \theta \boldsymbol{\nu} \cdot \boldsymbol{
\nabla} \theta \dd s
\label{contour_integral}
\end{equation}
Along the segments $x=0$, $x=L\cos\alpha$ and $y\to \infty$, we impose that
$\theta=\theta_{\mathrm{sing}}=\alpha_\infty$, where the far-field 
azimuthal angle is $\alpha_\infty=0$ for the $N^\perp$ texture and 
$\alpha_\infty=\pi/2$ for the $N^\parallel$ texture. 
We link $C_1$ to the $x=0$ and $x=L\cos\alpha$ segments by
circle arcs of small radii $\epsilon$. These circles take into 
account the presence of a defect core, and their radii are proportional to 
the nematic coherence length, which provides a microscopic cutoff for the 
continuum Frank-Oseen model. In these boundaries,
$\theta$ approaches, up to a constant,
the field of one isolated disclination line, which satisfies 
the following property 
$\boldsymbol{\nu}\perp \boldsymbol{\nabla}\theta$ as $\epsilon\to 0$. So we 
can neglect the contribution of the circle arcs to the contour integral.
Finally, $\theta=\alpha$ along the segment $C_1$ for strong anchoring 
conditions. Taking into account that $\oint_C \boldsymbol{\nu} \cdot 
\boldsymbol{\nabla} \theta \dd s=0$, we might rewrite Eq. 
(\ref{contour_integral}) as:
\begin{eqnarray}
\felastic&=&K\oint_C  (\theta-\alpha_\infty) 
\boldsymbol{\nu} \cdot \boldsymbol{\nabla} \theta \dd s\nonumber\\ &=&
K(\alpha-\alpha_\infty) \oint_{C_1} \boldsymbol{\nu} \cdot 
\boldsymbol{\nabla} \theta \dd s
\label{contour_integral2}
\end{eqnarray}
If $\theta$ is split into a the singular and non-singular contribution,
we see that the non-singular term, which can be expressed by Eq. 
(\ref{expansion}), leads to a contribution to $\felastic$ independent of $L$. 
On the other hand, the leading order contribution to $\felastic$ comes from 
the singular orientation field, namely the contribution close to the 
disclination lines and apex. If $s$ is the distance of a point of the
contour $C_1$ close to the wedge or apex, $\boldsymbol{\nu}\cdot 
\boldsymbol{\nabla} \theta_{\mathrm{sing}} \approx -I_1 / s$ 
or $-I_2 / s$, respectively. The leading order contribution to $\felastic$
comes from integration on $C_1$ as: 
\begin{equation}
\felastic \sim \mathcal{K}(\alpha)\ln L/\epsilon
\label{dominant_term}
\end{equation} 
where $\mathcal{K}$ is defined as:
\begin{equation}
\mathcal{K}(\alpha)=K(\alpha-\alpha_\infty)(I_2-I_1)=\begin{cases}
\frac{K\pi \alpha^2}{\left(\frac{\pi}{2}
\right)^2-\alpha^2} & \alpha<\frac{\pi}{4}\\ 
\\
K\pi\frac{\frac{\pi}{2}-\alpha}{\frac{\pi}{2}
+\alpha} & \alpha>\frac{\pi}{4}\\  
\end{cases}
\label{defkalpha}
\end{equation}
Expressions Eqs. (\ref{dominant_term}) and (\ref{defkalpha}) show that the 
equilibrium texture is $N^\perp$ for $\alpha<\pi/4$,
and $N^\parallel$ for larger values of $\alpha$, in agreement with our
LdG calculations and previous results reported in the literature 
\cite{Barbero_1980}.

\subsection{Analysis of the results within the LdG model} 

\begin{figure}[t]
\centerline{\includegraphics[width=.9\columnwidth]{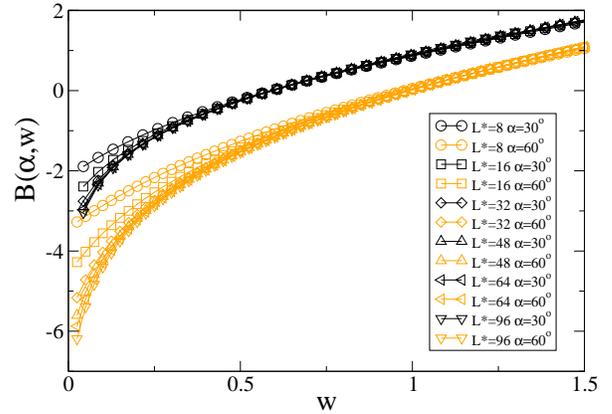}}
\caption{(Color online)
Plot of  $B(w,\alpha)$ as a function of the anchoring strength $w$ for
$\alpha=\pi/6$ and $\alpha=\pi/3$. The meaning of the symbols is the same as in
Fig.~\ref{fig4}.\label{fig6}}
\end{figure}

Finally, in order to check the accuracy of our approximation, we analyze
the results obtained with the LdG model by fitting the values of $\felastic$ 
obtained from minimization to an expression:
\begin{equation}
\felastic=\mathcal{K}(\alpha)\ln (L/\xi) + B(\alpha,w).
\label{elasticansatz}
\end{equation} 
where $\mathcal{K}(\alpha)$ is given by the expression Eq. (\ref{defkalpha})
and $B(\alpha,w)$ is expected not to depend explicitely on $L$ if $wL$ 
is large 
enough (i.e. under strong anchoring conditions). The resulting 
curves for different values of $L$ of $B(\alpha,w)$ as a function of $w$ 
for a fixed value of $\alpha$ collapse into a master curve (see 
Fig.\ref{fig6}). Deviations only appear for small values of $w\lesssim L^{-1}$, 
at which the strong anchoring condition breaks down. 

This result shows clearly that there is a contribution to the elastic part
of the surface free energy which scales logarithmically with the periodicity 
of the substrate. For large $L$ (small $q$) and $w\gg L^{-1}$, the surface
free energy density (i.e. per unit projected area on the $xz$ plane) of a 
nematic in the presence of a sawtooth substrate has the asymptotic
behaviour $\sigma\NW/\cos \alpha - (\mathcal{K}(\alpha)/2\pi)q\ln q + 
\mathcal{O}(q)$. The non-analytical contribution $-q\ln q$ introduces a  
slow decay of the surface free energy of the nematic with increasing $L$. 
This may help to explain the large deviations with respect to the Wenzel 
law observed for the wetting transition by nematic of a sawtoothed substrate 
in contact with the isotropic phase \cite{Patricio_Pham_Romero-Enrique}.  

\section{Conclusions}

In this paper we have analyzed the size dependence of the elastic 
contribution to the interfacial free energy density of a nematic
in presence of a sawtoothed substrate. The nucleation of non-half-integer 
disclination lines of the nematic director field on the apexes and wedges
of the substrate, predicted in the FO model, induce a non-analytical 
contribution which scales with the substrate periodicity wavenumber $q$
as $-q\ln q$ for small $q$. This has been confirmed by the numerical results in the full 
LdG model. The periodicity scaling of the elastic contribution to 
the surface free energy is different from that obtained for smooth surfaces, 
which scales linearly with $q$. 
Our arguments 
are not specific to this kind of substrate, and can be extended 
straightforwardly to any 
surface shape which shows ridges, cusps and similar singularities. On the 
other hand, the size-scaling of the elastic free energy has consequences for 
surface transitions such as wetting. Finally, the recent impressive advances in 
microfluidic technology and surface patterning open the posibility of 
an experimental verification of our predictions. 

\acknowledgments

The authors wish to thank Prof. M. M. Telo da Gama and Prof. A. O. Parry 
for enlightening discussions. We acknowledge the support from 
MICINN (Spain) through Grants No. HP2008-0028
and FIS2009-09326, and Junta de Andaluc\'{\i}a (Spain) through Grant No. 
P09-FQM-4938 (J.M.R.-E.), FCT (Portugal) through Grant No. 
SFRH/BPD/20325/2004 (C.-T.P.),
and Ac\c c\~ao Integrada Luso-Espanhola Ref. E 17/09 (P.P.).

%

\end{document}